\documentclass[11pt]{article}
\usepackage{amsmath, amssymb}
\newtheorem{lemma}{Lemma}[section]
\newtheorem{theorem}{Theorem}[section]

\newtheorem{definition}{Definition}[section]
\newtheorem{proposition}{Proposition}[section]
\def \AA  {{\cal A}}
\def \BB  {{\cal B}}

\def \EE  {{\cal E}}

\def \LL  {{\cal L}}

\def \OO  {{\cal O}}

\def \RR  {{\cal R}}

\def \TT  {{\cal T}}
\def \UU  {{\cal U}}

\def \BbR {{I\!\!R}}

\def \BbH {{I\!\!H}}
\begin{document}

\title{Mean field dynamics of fermions and the time-dependent Hartree-Fock equation}

%%
%% AUTHORS :
%%

\author{
%------------------
Claude BARDOS%
    \footnote{Laboratoire J.-L. Lions, Case 187, F75252 Paris Cedex 05
(bardos@math.jussieu.fr).
},\\
Fran\c cois GOLSE%
    \footnote{Laboratoire J.-L. Lions, Case 187, F75252 Paris Cedex 05,
(golse@math.jussieu.fr).
},\\
Alex D.\  GOTTLIEB %
\footnote{
Wolfgang Pauli Inst. c/o Inst.\ f.\ Mathematik, Univ. Wien, Strudlhofg.\ 4,
    A--1090 Wien, (alex@alexgottlieb.com).
} \\
and
Norbert J.\ MAUSER%
\footnote{
Wolfgang Pauli Inst. c/o  Inst.\ f.\ Mathematik, Univ. Wien, Strudlhofg.\ 4,
    A--1090 Wien, (mauser@courant.nyu.edu).
}
}
\maketitle

\begin{abstract}
The time-dependent Hartree-Fock equations are derived from the
$N$-body linear Schr\"odinger equation with mean-field scaling in
the limit $N\rightarrow \infty$ and for initial data that are like
Slater determinants. Only the case of bounded, symmetric
interaction potentials is treated in this work. We prove that, as
$N\rightarrow \infty$, the first partial trace of the $N$-body
density operator approaches the solution of the time-dependent
Hartree-Fock equations (in operator form) in the trace norm.
\end{abstract}


\smallskip
\noindent
{\bf Mathematical subject codes:} 35Q40, 35Q55, 47N50, 82C10, 82C22

%%%%%%%%%%%%%%%%%%%%%%%%%%%%%%%%%%%%%%%%%%%%%%%%%%%%%%%%%%%%%%%%%%%%%%%%

\section{Introduction}

In this article we consider the Hamiltonian dynamics of systems of
fermions and derive the time-dependent Hartree-Fock equation in
the mean field limit.   We follow the approach of Spohn, who
derived a mean field dynamical equation (the time-dependent
Hartree equation) for mean field systems of distinguishable
particles, remarking at the time that ``the convergence of the
mean field limit with statistics included is an open
problem"\cite{Sp}.

In Spohn's theory the initial $N$-body density operator $D_N$ is
assumed to be a product state $D^{\otimes N}$, i.e., the particles
are statistically independent and identically distributed. The
mean field limit is investigated in the Schr\"odinger picture,
where $D_N(t)$ obeys the von Neumann equation
\begin{equation}
i\hbar  \frac{d}{dt}D_N(t) \ = \ \sum_{1 \le j \le N} \big[ L_j, D_N(t)
\big] \ + \  \frac{1}{N} \sum_{1 \le i<j \le
N} \left[ V_{ij},D_N(t) \right]
\label{Schrodinger equation for the Intro}
\end{equation}
with $V_{ij}$ denoting the two-body potential $V$ acting between
the $i^{th}$ and $j^{th}$ particles and $[\ ,\ ]$ denoting the
commutator.  The limit as $N \longrightarrow \infty$ of the
$n$-body density operator $D_{N:n}(t)$ is shown to converge to
$D(t)^{\otimes n}$, where $D(t)$ obeys a time-dependent Hartree
equation.   (The subscript $_{:n}$ appearing in $D_{N:n}$ is our
notation for the $n^{th}$ partial trace, defined in equation
(\ref{partial trace}) below.) Spohn's ideas have been generalized
to open systems in \cite{Alicki}.  There are other theories of
quantum mean field dynamics, e.g., the algebraic theory of
\cite{DW}, but to our knowledge the problem of including quantum
statistical effects remains unsolved.

The problem is that Fermi-Dirac or Bose-Einstein statistics
constrain the possible initial condition of (\ref{Schrodinger
equation for the Intro}) to have the appropriate symmetry, which
is typically inconsistent with the product form $D^{\otimes N}$.
An $N$-body density operator with Fermi-Dirac symmetry can never
have the form $D^{\otimes N}$ and a Bose-Einstein density operator
can only have the form $D^{\otimes N}$ if $D$ is a pure state
(i.e., if the system of bosons is in a condensed state).  The
remedy for this problem, for fermions, is to replace the
hypothesis that the initial state be a product state with a
hypothesis that is consistent with Fermi-Dirac statistics, e.g.,
that the initial states are Slater determinants.

The role of the factorization hypothesis $D_N(0)=D^{\otimes N}$ is
to permit the closure of the BBGKY hierarchy by setting the
two-body state $D_{N:2}$ equal to $D\otimes D$. Closing the
hierarchy this way results in the time-dependent Hartree equation.
This kind of closure hypothesis is implicit in the {\it
Stosszahlansatz} that leads to Boltzmann's kinetic equation for
gases \cite{Boltzmann}.  Kac noted that, for Boltzmann's equation,
the factorization $f_{N:2} = f\otimes f$ is only realized in the
limit $N \longrightarrow \infty$, and he called this behavior the
{\it Boltzmann property} \cite{Kac55,Kac}. Later authors
\cite{McK66,Gru,Szn} developed Kac's ideas; what is now called the
{\it propagation of chaos} is an important tool in rigorous
kinetic theory \cite{Sznitman, Meleard, Thesis}. We have noted
that Boltzmann's closure Ansatz is inconsistent with the Pauli
Exclusion Principle, and needs to be replaced by another closure
Ansatz when the particles are fermions. The novelty of our
approach consists in replacing the condition of asymptotic
independence of the particles by a condition that describes the
correlations of Slater determinants. This condition, called {\it
Slater closure} is defined in Definition~\ref{closure} below.

Assuming that $\{D_N(0)\}$ is a sequence of initial states for
(\ref{Schrodinger equation for the Intro}) that has Slater closure,
we can prove that $\{D_N(t)\}$ has Slater closure for all $t>0$.
This phenomenon could be called the {\it propagation of Slater closure}
because it is like the ``propagation of chaos" mentioned above.
Since $\{D_N(t)\}$ has Slater closure, the two-body density operator
$D_{N:2}(t)$ is approximately equal to
$(D_{N:1}(t) \otimes D_{N:1}(t)) \Sigma_2$ when $N$ is large,
where $\Sigma_2$ is the two-body operator
defined by
\[
    \Sigma_2 (x\otimes y) \ = \ x\otimes y - y \otimes x \ .
\]
Substituting $(D_{N:1}(t) \otimes D_{N:1}(t)) \Sigma_2$ for $D_{N:2}(t)$
in the BBGKY hierarchy leads one to conjecture that,
when $N$ is large,
the single-body density operator should nearly obey
the time-dependent Hartree-Fock (TDHF) equation
\begin{eqnarray*}
          i\hbar  \frac{d}{dt}F(t)
          & = &
           \left[ L, \ F(t)\right]   \ + \     \left[ V,\ (F(t)\otimes
F(t)) \Sigma_2 \right]_{:1}  \\
          F(0)
          & = &  D_{N:1}(0) .
\end{eqnarray*}
Theorem~\ref{Main result} confirms this conjecture.

Theorem~\ref{Main result} states that the distance in the trace
norm between $D_{N:1}(t)$ and the corresponding solution $F(t)$ of
the TDHF equation tends to $0$ as $N$ tends to infinity.   The
trace norms of $D_{N:1}(t)$ and $F(t)$ are separately equal to
$1$, so it is significant that their difference $D_{N:1}(t)-F(t)$
converges to $0$ in the trace norm.  A crucial detail of the proof
is Lemma~\ref{lemma1}, which states that the {\it operator} norm
of $D_{N:1}$ tends to $0$ if $\{D_N \}$ has Slater closure.  Much
of the rest of the proof lies in bounding the {\it trace norm} of
$D_{N:1}(t)-F(t)$ by an expression involving the {\it operator
norm} of $D_{N:1}(0)$.

The use of the trace norm to measure the distance between two
density operators is quite natural. A density operator $D$
corresponds to a quantum state through the assignment $ B \mapsto
\hbox{Tr}(DB)$ of expectation values to bounded observables $B$.
Thus, two density operators $D$ and $D'$ are within $\varepsilon$
of one another in the trace norm if and only if they correspond to
quantum states that give expectations differing by no more than
$\varepsilon$ for all observables $B$ with $\|B\| \le 1$.

In this article, we assume that the two-body potential $V$ is a
bounded operator.  Recent work on the time-dependent
Schr\"odinger-Poisson equation \cite{BEGMY,EY} suggests that it
may be possible to prove a theorem similar to our
Theorem~\ref{Main result} when $V$ is the Coulomb potential.  This
work shall be published in a separate paper.

The rest of this article is organized as follows: the next
section discusses fermionic density operators and defines Slater
closure.  The N-particle Hamiltonian and the associated
time-dependendent Hartree-Fock equation are described in
Section~3. That section concludes with the statement of our main
result, Theorem~\ref{Main result}, whose proof spans Sections~4,
5, and 6.  Section~7 is an appendix relating the operator form
of the TDHF equation, used throughout this paper, to the
formulation of the TDHF equation as a coupled system of wave
equations, which may be more familiar to some readers.

\section{Fermionic density operators and Slater closure}

Let $\BbH$ be a Hilbert space, supposed to be the space of wavefunctions for a
certain type of quantum system (a ``component" or ``particle").  Then the
Hilbert space of wavefunctions for a system consisting of $N$ distinguishable
components or particles of that type is
$\BbH_N = \BbH^{\otimes N}$.   If the components are not distinguishable, but
obey
Fermi-Dirac statistics, then the appropriate Hilbert space of wavefunctions
is the antisymmetric subspace $\AA_N \subset \BbH_N$.
 To define this subspace, it is convenient first to define unitary {\it
transposition} and {\it permutation} operators
 on $\BbH_N$.
  The transposition operator $U_{(ij)}$ is defined by
extending the following isometry defined on {\it simple tensors}
\[
U_{(ij)}(x_1\otimes x_2 \otimes \cdots x_i \cdots x_j \cdots \otimes x_N)
\ = \
 x_1\otimes x_2 \otimes \cdots x_j \cdots x_i \cdots \otimes x_N
\]
to all of $\BbH_N$.  For any $\pi$ in the group $\Pi_N$ of
permutations of $\{1,2,\ldots,N\}$, one may define the permutation operator
$ U_{\pi} $ as $ U_{(i_k j_k)} \cdots U_{(i_2 j_2)} U_{(i_1 j_1)}$,
where $(i_k j_k)
\cdots (i_2 j_2)(i_1 j_1)$ is any product of transpositions that equals $\pi$.

The antisymmetric subspace may now be defined as
\[
           \AA_N      \ = \
           \left\{ \psi \in \BbH_N \ : \    U_{\pi}\psi =
\hbox{sgn}(\pi)\psi \quad  \forall \pi \in \Pi_N \right\}      .
\]
  One may verify that
$$P_{\AA_N} \ =
\  \frac{1}{N!} \sum_{\pi \in \Pi_N} \hbox{sgn}(\pi) U_{\pi}$$
is the orthogonal projector whose range is $\AA_N$.

The {\it pure states} of an $N$-fermion system correspond to the
orthogonal projectors $P_{\psi}$ onto one-dimensional subspaces of $\AA_N$.
That is, a pure state is given by
$$
 P_{\psi}(\phi) \ = \ \langle \psi, \phi \rangle \psi
$$
for some $\psi \in \AA_N$ of unit length.  The {\it statistical
states} of the $N$-fermion system are the positive trace class
operators or {\it density operators} $D$ on $\AA_N$ of trace $1$.
These can be identified with density operators $D$ on all of
$\BbH_N$ whose eigenvectors lie in $\AA_N$, i.e., such that
$$
  D \ = \ \sum_{i=1}^{\infty} \lambda_i P_{\psi_i}
$$
for some orthonormal system $\{\psi_i\}$ in $\AA_N$ and a family
of positive numbers $\lambda_i$ that sum to $1$.  It follows that
these {\it fermionic densities} are those density operators that
satisfy
\begin{equation}
\label{antisymmetric}
D U_{\pi} \ = \ U_{\pi} D \ = \ \hbox{sgn}(\pi) D
\qquad \qquad
\forall \pi \in \Pi_N.
\end{equation}
If a density operator $D$ on $\BbH_N$ commutes with every
permutation operator $U_{\pi}$ then it is {\it symmetric}.  In
particular, fermionic densities are symmetric by
(\ref{antisymmetric}).

If $\{e_j\}_{j \in J}$ is an orthonormal basis of $\BbH$ then
$$
\left\{ e_{j_1}\otimes e_{j_2} \otimes \cdots \otimes e_{j_N}: \
j_1,j_2,\ldots,j_N \in J  \right\}$$ is an orthonormal basis of
$\BbH_N$.   Since $\AA_N$ is the range of  $P_{\AA_N}$ and since
$$
P_{\AA_N}(e_{j_1}\otimes e_{j_2} \otimes \cdots \otimes e_{j_N}) \
= \ 0$$ unless all of the indices $j_i$ are distinct, the set
$$   \left\{
P_{\AA_N}(e_{j_1}\otimes e_{j_2} \otimes \cdots \otimes e_{j_N})\!
: \ j_1, j_2 , \ldots, j_n \ \hbox{all distinct}\right\}
$$
is a spanning set for $\AA_N$.  In fact it is an orthogonal basis
for $\AA_N$, each vector having norm $1/\sqrt{N!}$.  Vectors of
the form  $\sqrt{N!}\  P_{\AA_N}(e_{j_1}\otimes e_{j_2} \otimes
\cdots \otimes e_{j_N})$ are known as {\it Slater determinants}.

The trace class operators on a Hilbert space $\BbH$ form a Banach
space $\TT(\BbH)$ with the norm $\Vert T\Vert_{tr}  =
\hbox{Tr}(|T|)$.  The important inequality
\begin{equation}
\Vert T B \Vert_{tr} \ \le \ \Vert T \Vert_{tr} \Vert B
\Vert
\label{basic inequality}
\end{equation}
holds whenever $B$ is a bounded operator of norm $\Vert B \Vert$
and $T \in \TT(\BbH)$.
It is this basic inequality that will produce our key estimates.

For $n \le N$, the $n^{th}$
{\it partial trace} is a contraction from
$\TT(\BbH^{\otimes N})$ onto $\TT(\BbH^{\otimes n})$.  The $n^{th}$
partial trace of $T$ will be denoted $T_{:n}$, and may be defined
as follows:  Let $\OO$ be any orthonormal basis of $\BbH$. If $T
\in \TT(\BbH^{\otimes N})$ and $n < N$ then
\begin{eqnarray}
\left<T_{:n}(w),x \right> \ =&\nonumber
\\
\sum_{z_1,\ldots,z_{N-n} \in \ \OO}
&\Big<\ T( w \otimes z_1 \otimes \cdots \otimes z_{N-n}),\ x
\otimes z_1 \otimes \cdots \otimes z_{N-n} \Big>
\label{partial trace}
\end{eqnarray}
for any $w,x \in \BbH^{\otimes n}$.   If a trace class operator $T
\in \TT(\BbH^{\otimes N})$ satisfies (\ref{antisymmetric}) then so
does $T_{:n}$, i.e., the partial trace defines a positive
contraction from $\TT(\BbH^{\otimes N})$ to $\TT(\BbH^{\otimes
n})$ that carries fermionic densities to fermionic densities.

In the following definition, and throughout this article, we use
the superscript $^{\otimes n}$ to denote the $n^{th}$ tensor power
of an operator, and we use the notation $\Sigma_n$ for
$n!P_{\AA_n}$, i.e.,
\[
  \Sigma_n \ = \ \sum_{\pi \in \Pi_n} \hbox{sgn}(\pi)
U_{\pi} \ .
\]
 The $n^{th}$
{\it tensor power} of an operator $A$ on $\BbH$ is the operator
$A^{\otimes n}$ on $\BbH_n$ defined on simple tensors by
\[
     A^{\otimes n}(x_1\otimes x_2 \otimes \cdots \otimes x_n)
     \ = \
     Ax_1 \otimes Ax_2 \otimes \cdots \otimes Ax_n\ .
\]
\begin{definition}
\label{closure}
For each $N$, let $D_N$ be a symmetric density operator on $\BbH_N$.
 The sequence $\left\{D_N\right\}$ {\bf has Slater closure} if, for each fixed
$n$,
\[
              \lim_{N \rightarrow \infty}    \big\|      D_{N:n} \ - \
D_{N:1}^{\otimes n} \Sigma_n  \big\|_{tr} \ = \ 0.
\]
\end{definition}

This terminology is motivated by the observation that, if $\Psi_N$
is a Slater determinant in $\AA_N$ and $P_{\Psi_N}$ denotes the
orthogonal projector onto the span of $\Psi_N$ then
\begin{equation}
 \left(P_{\Psi_N}\right)_{:n} \ = \ \frac{N^n(N-n)!}{N!}
  \left(P_{\Psi_N}\right)_{:1}^{\otimes n} \sum_{\pi \in \Pi_n}
\hbox{sgn}(\pi)U_{\pi} \ ,
\label{SlaterFormula}
\end{equation}
for this implies the following:

\begin{proposition}
\label{Slater case}
For each $N$ let $\Psi_N$ be a Slater determinant in $\AA_N$, and let
$P_{\Psi_N}$ denote the orthoprojector onto the span of $\Psi_N$.
Then $\left\{  P_{\Psi_N}  \right\}$ has Slater closure.
\end{proposition}

%%%%%%%%%%%%%%%%%%%%%%%%%%%%%%%%%%%%%%%%%%%%%%%%%%%%%%%%%%%%%%%%%%%%%%%%%
%
%
%%%%%%%%%%%%%%%%%%%%%%%%%%%%%%%%%%%%%%%%%%%%%%%%%%%%%%%%%%%%%%%%%%%%%%%%%

\section{The time-dependent Hartree-Fock equation}

We are going to prove that, in the mean field limit,  the
time-dependent Hartree-Fock equation describes the time-evolution
of the single-particle state in systems of fermions.   We state
our theorem in this section and go on to prove it in the three
subsequent sections.

First we describe the $N$-particle Hamiltonian.  Let $L^{(N)}$ be
a self-adjoint operator on $\BbH$, where $L^{(N)}$ may depend on
$N$ in an arbitrary manner. The free motion of the $j^{th}$
particle is governed by
\[
         L^{(N)}_j = I^{\otimes j-1}\otimes L^{(N)} \otimes I^{\otimes N-j},
\]
where $I$ denotes the identity operator on $\BbH$.   The
interaction between the particles has the form $1/N$ times the sum
over pairs of distinct particles of a two-body potential $V$. Let
$V$ be a bounded Hermitian operator on $\BbH \otimes \BbH$ that
commutes with the transposition operator $U_{(12)}$.   Define the
operator $V_{12}$ on $\BbH_N$ by
$$
V_{12}\left( x_1\otimes x_2 \otimes \cdots \otimes x_N \right)
\ = \
V(x_1\otimes x_2) \otimes x_3 \otimes \cdots \otimes x_N
$$
and for each $1 \le i < j \le N$ define $V_{ij}= U^*_{\pi} V_{12}
U_{\pi}$ where $\pi$ is any permutation with $\pi(i)=1$ and $\pi(j)=2$.  Let
\begin{equation}
\label{Ham} H_N \ = \ \sum_{1 \le j \le N}L^{(N)}_j \ + \
\frac{1}{N} \sum_{1 \le i<j \le N} V_{ij}
\end{equation}
be the {\it N-particle Hamiltonian} operator on $\BbH_N$. The von
Neumann equation for the $N$-particle density operator $D_N(t)$ is
\begin{equation}
i\hbar  \frac{d}{dt}D_N(t) \ = \ \sum_{1 \le j \le N} \big[
L^{(N)}_j, D_N(t) \big] \ + \  \frac{1}{N} \sum_{1 \le i<j \le N}
\left[ V_{ij},D_N(t) \right] \label{N-particle Schrodinger}
\end{equation}
and has the solution
\begin{equation}
 e^{-iH_Nt/\hbar}D_N(0)e^{iH_Nt/\hbar}.
\label{N-particle solution}
\end{equation}

Next we define the time-dependent Hartree-Fock equation. Let $L^{(N)}$ and
$V$ be as above, and let $U_{(12)}$ denote the transposition operator on
$\BbH\otimes \BbH$. The time-dependent Hartree-Fock (TDHF) equation for a
density operator $F(t)$ on $\BbH$ is
\begin{eqnarray}
          i\hbar  \frac{d}{dt}F(t)
          & = &
          \big[ L^{(N)}, \ F(t)\big]   \ + \     \left[ V,\ F_2^-(t) \right]_{:1} \nonumber \\
          F_2^-(t)
      & = &
          (F(t)\otimes F(t)) \left(I - U_{(12)}\right)
      \ = \
      F(t)^{\otimes 2}  \Sigma_2
\label{TDHF}
\end{eqnarray}
(the subscript $_{:1}$ on the last commutator denotes partial contraction).
Following \cite{Bove}, we define a {\bf mild solution} of equation (\ref{TDHF})
to be a continuous function $t\mapsto F(t)$ from $[0,\infty)$ to the real Banach
space of Hermitian trace class operators such that
\[
F(t)=
e^{-i\frac{t}{\hbar}L^{(N)}}F(0)e^{+i\frac{t}{\hbar}L^{(N)}}
-\frac{i}{\hbar}\int_0^te^{-i\frac{t-s}{\hbar}L^{(N)}}
    [V,F_2^-(s)]_{:1}e^{+i\frac{t-s}{\hbar}L^{(N)}}ds
\]
for all $t\ge 0$. The results proved in \cite{Bove} show that
(\ref{TDHF}) has a global mild solution\footnote{The solution
obtained in Theorem 4.2 of \cite{Bove} is indeed defined for
all positive times because the nonlinearity of TDHF satisfies
condition (4.1) of \cite{Bove} --- see Proposition 3.5 there.}
for each Hermitian trace class initial data $F(0)$. Furthermore,
\begin{equation}
\label{nonlinearUnitary}
F(t) =
U^*F(0) U
\end{equation}
for some unitary operator depending on
$t$ and $F(0)$. In particular, the operator norm of $F(t)$
is constant.

The relationship between the $N$-particle system and the TDHF
equation is the subject of our main theorem. Recall the definition
(\ref{closure}) of Slater closure.

\begin{theorem}
\label{Main result}
For each $N$, let $D_N(t)$ the solution to (\ref{N-particle Schrodinger})
whose initial value $D_N(0)$ is a symmetric density.
Let $F^{(N)}(t)$ be the mild solution of the TDHF equation (\ref{TDHF})
whose initial value is $F^{(N)}(0)=D_{N:1}(0)$.

If $\left\{ D_N(0) \right\}$ has Slater closure then $\left\{ D_N(t) \right\}$
has Slater closure and
\[
             \lim\limits_{N \rightarrow \infty} \big\| D_{N:1}(t) - F^{(N)}(t)
\big\|_{tr} \ = \  0
\]
for all $t > 0$.
\end{theorem}

%%%%%%%%%%%%%%%%%%%%%%%%%%%%%%%%%%%%%%%%%%%%%%%%%%%%%%

\section{Two hierarchies, and their difference}

Consider the $N$-particle von Neumann equation (\ref{N-particle
Schrodinger}). {From} now on we will suppose that the initial
$N$-particle density operator $D_N(0)$ is symmetric, i.e., that
\[
                    U_{\pi}^* D_N(0) U_{\pi} =  D_N(0)
\] for all $\pi \in \Pi_N$.  (Recall that, in particular, fermionic densities
are symmetric.) The symmetry of the Hamiltonian (\ref{Ham})
ensures that $D_N(t)$ remains symmetric for all $t$. {From}
(\ref{N-particle Schrodinger}) and the symmetry of $D_N(t)$, it
follows that the partial trace $D_{N:n}(t)$ satisfies\footnote{In
the proof of Theorem \ref{Main result}, we proceed as if $D_N$ was
a classical solution of equation (\ref{N-particle Schrodinger});
the case of general initial data as in the statement of the Theorem
follows from a density argument and the unitarity of the group
generated by the $N$-body Schr\"odinger operator.}

\begin{eqnarray}
           i\hbar  \frac{d}{dt}D_{N:n}(t)
           & = &
          \sum_{1 \le j \le n} \big[ L^{(N)}_j,D_{N:n}(t)  \big]     \ + \
          \frac{1}{N} \sum_{1 \le i<j \le n} \left[ V_{ij},D_{N:n}(t)
           \right]
                       \nonumber        \\
           &   &
           + \ \frac{N-n}{N} \sum_{1 \le i \le n} \left[
            V_{i,n+1},D_{N:n+1}(t)\right]_{:n}
\label{N-particle hierarchy}
\end{eqnarray}
The system of equations (\ref{N-particle hierarchy}) for
$D_{N:1},D_{N:2},\ldots,D_{N:N-1}$ together with the equation
(\ref{N-particle Schrodinger}) for $D_N$
is called the {\it N-particle hierarchy}.
For our estimates later on, it is convenient to rewrite the equations
(\ref{N-particle hierarchy}) of the hierarchy as
\begin{eqnarray}
           i\hbar  \frac{d}{dt}D_{N:n}(t)
           & = &
          \LL^{(N)}_n( D_{N:n}(t) )    \ + \   \sum_{1 \le i \le n} \left[
            V_{i,n+1},D_{N:n+1}(t)\right]_{:n}   \nonumber \\
           &   &
            + \  \EE_n(t,N,D_N(0))
\label{N-particle hierarchy rewritten}
\end{eqnarray}
with
\begin{eqnarray}
          \LL^{(N)}_n(\ \cdot \ )
           & = &
          \sum_{1 \le j \le n} \big[ L^{(N)}_j,   \   \cdot \  \big]
\nonumber   \\
            \EE_n(t,N,D_N(0))
           & = &
           \frac{1}{N} \sum_{1 \le i<j \le n} \left[ V_{ij}, D_{N:n}(t)
\right]
\nonumber   \\
           &   &
           - \ \frac{n}{N}\sum_{1 \le i \le n} \left[
V_{i,n+1},D_{N:n+1}(t)\right]_{:n} \ .  \label{N-particle hierarchy stuff}
\end{eqnarray}

Next we describe another hierarchy, built from ``the bottom up" out of
solutions to the TDHF equation, in contrast to the hierarchy we have just
considered, which is built from ``the top down" starting with solutions to
(\ref{N-particle Schrodinger}).
If $F$ is a trace class operator, define $F^-_1 = F$ and
\[
         F^-_n \ = \  F^{\otimes n} \Sigma_n
\]
for $n >1$.  When $F$ depends on $t$ we write $F^-_n(t)$ instead
of $F(t)^-_n$.   The notation $F^-_2(t)$ has already been used in
the TDHF equation (\ref{TDHF}).

\begin{proposition}\label{Prop1}
If $F$ is a classical solution \footnote{Consistently with the
previous footnote, we are proceeding as if $F_N$ was a classical
solution of (\ref{TDHF}); the case of general initial data as in
the statement of the Theorem is recovered by a density argument
and the continuous dependence on initial data of the mild solution
to (\ref{TDHF}): see Theorem 4.1 of \cite{Bove}.} of the TDHF
equation (\ref{TDHF}) then
\[
          i\hbar  \frac{d}{dt}F^-_n(t) \ = \
            \sum_{j=1}^n  \big[ L^{(N)}_j, F^-_n(t) \big]   \  + \
           \sum_{j=1}^n  \left[ V_{j,n+1},\ F^-_{n+1}(t)\right]_{:n} \ +\
\RR_n(F(t))
\]
where $\RR_n$ is defined on trace class operators by $\RR_1(X) = {\bf 0}$ (the zero
operator) and
\begin{equation}
\label{Rn}
      \RR_n(X) \ = \   \sum_{j=1}^n \Big[V_{j,n+1}, \ X^{\otimes n + 1}
        \sum_{k \ne j}
          U_{(k,n+1)}\Big]_{:n}\Sigma_n
\end{equation}
for $n > 1$.
\end{proposition}

\noindent {\bf Proof}:  \qquad
For any trace class operator $X$,
\begin{eqnarray}
        &&\sum_{j=1}^n\big[ V_{j,n+1}, \ X^-_{n+1}\big]_{:n}
        \nonumber\\
        & = &
        \sum_{j=1}^n \Big[ V_{j,n+1}, \ X^{\otimes n + 1} \Big( I -
\sum_{k=1}^n U_{(k,n+1)} \Big)
           \Sigma_n \otimes I_{\BB(\BbH)}     \Big]_{:n}
        \label{FrancoisWantsNumberHere} \\
        & = &
        \sum_{j=1}^n \Big[ V_{j,n+1}, \ X^{\otimes n + 1} \big( I - \sum_{k=1}^n
U_{(k,n+1)} \big)
              \Big]_{:n} \Sigma_n \ .
        \nonumber
\end{eqnarray}
The first equality in (\ref{FrancoisWantsNumberHere}) holds
because
\[
\Sigma_{n+1}=\Big( I - \sum_{k=1}^n U_{(k,n+1)} \Big)
           \Sigma_n \otimes I_{\BB(\BbH)} ,
\]
and the second equality in (\ref{FrancoisWantsNumberHere}) holds
because $\Sigma_n\otimes I_{\BB(\BbH)}$ commutes with
$\sum\limits_{j=1}^n V_{j,n+1}$. {From} the TDHF equation
(\ref{TDHF}) we calculate that
\begin{eqnarray}
         && i\hbar \frac{d}{dt}F^-_n(t)
           =
          i\hbar \frac{d}{dt} F(t)^{\otimes n}\Sigma_n  \nonumber\\
          & = &
          i\hbar \Big\{
          \sum_{j=1}^n F(t)^{\otimes j-1} \otimes \frac{d}{dt}F(t) \otimes
F(t)^{\otimes n-j} \Big\}     \Sigma_n
        \nonumber \\
        & = &
         \sum_{j=1}^n \left\{
            \big[ L^{(N)}_j, F(t)^{\otimes n} \big]         +
            \left[ V_{j,n+1}, \ F(t)^{\otimes n + 1}\left(I -
U_{(j,n+1)}\right) \right]_{:n}
           \right\} \Sigma_n
        \nonumber  \\
        & = &
            \sum_{j=1}^n \big[ L^{(N)}_j, F^-_n(t) \big] \ + \  \RR_n(F(t))
        \label{AlexPutNumberHere} \\
        &   & + \
            \sum_{j=1}^n \Big[ V_{j,n+1}, \ F(t)^{\otimes n + 1} \Big( I -
\sum_{k=1}^n U_{(k,n+1)} \Big)
              \Big]_{:n} \Sigma_n \ .
        \nonumber
\end{eqnarray}
By the identity (\ref{FrancoisWantsNumberHere}), the last sum in
(\ref{AlexPutNumberHere}) equals $\sum\limits_{j=1}^n \big[
V_{j,n+1}, \
        F^-_{n+1}(t)\big]_{:n}$, proving the proposition.
\hfill $\square$

Now let $D_N(t)$ be a solution of the N-particle von Neumann
equation (\ref{N-particle Schrodinger}) and let  $F(t)$ be a
solution of the TDHF equation (\ref{TDHF}). For $1 \le n \le N$
define the $n^{th}$ {\it difference}
\begin{equation}
           E_{N,n}(t)   \ = \  D_{N:n}(t)   - F^-_n(t)  .
\label{difference}
\end{equation}
{From} the N-particle hierarchy equations (\ref{N-particle hierarchy
rewritten}) and (\ref{N-particle hierarchy stuff}) and
Proposition~\ref{Prop1}, it follows that
\begin{eqnarray}
          i\hbar  \frac{d}{dt} E_{N,n}(t)
         & = &
          \LL^{(N)}_n( E_{N,n}(t))  \  + \
           \sum_{j=1}^n  \left[ V_{j,n+1},\ E_{N,n+1}(t)\right]_{:n}
\nonumber \\
         &   &
         + \   \EE_n(t,N,D_N(0))  \ - \   \RR_n(F(t))
\label{Hierarchy of differences}
\end{eqnarray}
for $n=1,2,\ldots,N-1$.
The characters $\EE$ and $\RR$ were chosen to evoke the words ``error" and
``remainder."  Indeed, in the next section we find bounds on these error terms
under conditions on $D_N(0)$ and $F(0)$.    The rest of this section is
devoted to showing
how such bounds lead to an upper bound on the differences $E_{N,n}(t)$.

To this end, let us define
\begin{equation}
            \hbox{Err}(t,N, n)  \ = \  \EE_n(t,N,D_N(0)) \ - \   \RR_n(F(t)).
\label{err}
\end{equation}
Let $U^{(N)}_{n,t}$ denote the unitary operator
$\exp\big(\frac{it}{\hbar}\sum_{j=1}^n L_j^{(N)}\big) $ on
$\BbH_n$ and define isometries $\UU^{(N)}_{n,t}$ on the trace
class operators by
\[
               \UU^{(N)}_{n,t}(\  \cdot \ )  \ = \  e^{\frac{it}{\hbar}
\LL^{(N)}_n}(\  \cdot \ ) \ = \     U^{(N)}_{n,t}(\  \cdot \ )
U^{(N)}_{n,-t} \  .
\]
Then  $Z_{N,n}(t) =  \UU^{(N)}_{n,t}(E_{N,n}(t))$ has the same
trace norm as $E_{N,n}(t)$ and satisfies
\begin{equation}
\label{Z hierarchy}
              \frac{d}{dt}  Z_{N,n}(t) \ = \      -\frac{i}{\hbar}
\sum_{j=1}^n  \left[ V_{j,n+1},\ Z_{N,n+1}(t)\right]_{:n}  -\frac{i}{\hbar}
\UU^{(N)}_{n,t}\hbox{Err}(t,N, n)
\end{equation}
for $n=1,2,\ldots,N-1$.    {From} (\ref{Z hierarchy}) it follows that
\begin{eqnarray*}
            \big\| E_{N,n}(t) \big\|_{tr}   & = &
            \big\|  Z_{N,n}(t) \big\|_{tr}
            \\
            & \le &
            \big\|  Z_{N,n}(0) \big\|_{tr} \ + \  \frac{2\|V\|n}{\hbar}
\int_0^t \big\|  Z_{N,n+1}(s) \big\|_{tr}ds\\
           & + & \   \frac{1}{\hbar}  \int_0^t
           \big\|\UU^{(N)}_{n,t}(\hbox{Err}(s,N,n))\big\|_{tr}ds
\end{eqnarray*}
for $n=1,2,\ldots,N-1$.  Recalling that $\|Z_{N,n+1}(s)\|_{tr}=\|E_{N,n+1}(s)\|_{tr}$
and that $\|\UU^{(N)}_{n,t}(\hbox{Err}(s,N,n))\big\|_{tr}=\|\hbox{Err}(s,N,n)\|_{tr}$,
the preceding inequality becomes
\begin{equation}
        \big\| E_{N,n}(t) \big\|_{tr} \ \le \
        \varepsilon(N,n,t) \ + \  \frac{2\|V\|n}{\hbar}
            \int_0^t \big\|E_{N,n+1}(s) \big\|_{tr}ds
\label{iterate me}
\end{equation}
if we define
\begin{equation}
\varepsilon(N,n,t)     \ = \   \big\|  E_{N,n}(0) \big\|_{tr} \ +
    \  \frac{1}{\hbar}  \int_0^t \big\| \hbox{Err}(s,N,n) \big\|_{tr}ds.
\label{eps}
\end{equation}
Beginning from (\ref{iterate me}) and iterating the inequality $m$ times (for
some $m \le N-n-1$)
we obtain our desired bound on the trace norm of $E_{N,n}(t)$:
\begin{eqnarray}
       \big\| E_{N,n}(t) \big\|_{tr}
       & \le &
      \sum_{k = 0}^{m}  \binom{n+k-1}{n-1} \left(\frac{2\|V\|t }{\hbar}
\right) ^k \varepsilon(N,n+k,t)  \nonumber \\
       & + &
           \binom{n+m-1}{n-1} \left(\frac{2\|V\|t }{\hbar} \right) ^{m}
\sup_{s \in [0,t]}\Big\{ \big\|  E_{N,n+m+1}(s) \big\|_{tr}  \Big\} .
       \nonumber \\
\label{Bardos inequality}
\end{eqnarray}

%%%%%%%%%%%%%%%%%%%%%%%%%%%%%%%%%%%%%%%%%%%%%%%%%%%%%%%%%%%%%%%%%%%%%

\section{Error estimates}

In this section we collect the error estimates that will be used
to prove Theorem~\ref{Main result}.

If $D_N(0)$ is a density operator then the solution $D_N(t)$ of
the N-particle von Neumann equation (\ref{N-particle Schrodinger})
is a density operator for all $t>0$, and it is clear from
(\ref{N-particle hierarchy stuff}) that
\begin{equation}
          \|  \EE_n(t,N,D_N(0)) \|_{tr}
          \   \le  \
          3\frac{n^2}{N} \| V \|  \
\label{estimates for N-particle hierarchy stuff}
\end{equation}
for all $t$.

\begin{lemma}
\label{lemma1}
If  $\left\{D_N\right\}$ has Slater closure then
\[
          \lim\limits_{N \rightarrow \infty} \|D_{N:1}\| = 0.
\]
\end{lemma}

 \noindent {\bf Proof}:  \qquad    The trace norm of $D_{N:1}^2$ equals the
sum of the squares of the eigenvalues of $D_{N:1}$. Since the
operator norm of $D_{N:1}$ equals its largest eigenvalue, it
follows that $\big\| D_{N:1} \big\| \le \left\| D_{N:1}^2
\right\|_{tr}^{1/2}$.  But $D_{N:1}^2 = \left\{ D_{N:1}^{\otimes
2}U_{(12)}\right\}_{:1}$, whence
\[
      \left\| D_{N:1}^2 \right\|_{tr}  \ = \    \Big\| \Big\{D_{N:2} -
D_{N:1}^{\otimes 2}(I - U_{(12)}) \Big\}_{:1} \Big\|_{tr}
      \ \le \    \Big\| D_{N:2} - D_{N:1}^{\otimes
2}\Sigma_2\Big\|_{tr}.
\]
The Slater closure of $\left\{D_N\right\}$ implies that the
right-hand side of the preceding inequality tends to $0$ as $N
\longrightarrow \infty$. \hfill $\square$

\begin{lemma}
 If $F$ is a density operator then $\big\| F_n^- \|_{tr} \le 1$ for all $n$.
\label{estimate for Fn-}
\end{lemma}

\noindent {\bf Proof}:  \qquad   Since $ \Sigma_n =\big(\Sigma_n\big)^* =
\frac{1}{n!} \big(\Sigma_n \big)^2   $ commutes with $F^{\otimes n}$, it
follows that
$
               F_n^- = F^{\otimes n}\Sigma_n = \frac{1}{n!}
\Sigma_n \left( F^{\otimes n} \right) \Sigma_n $ is a nonnegative
operator.  Thus, the trace norm of $F_n^-$ equals its trace. This
trace is
\[
                 \sum_{j_1,\ldots,j_n \in J} \Big< e_{j_1} \otimes \cdots
\otimes e_{j_n},\ F^{\otimes n} \Sigma_n(e_{j_1} \otimes \cdots
\otimes e_{j_n})\Big>
\]
where $\{e_j\}_{j \in J}$ is an orthonormal basis for $\BbH$ consisting of
eigenvectors of $F$.  This sum may be taken over distinct indices $j_1,
\ldots,j_n \in J$, since $\Sigma_n$ annihilates all tensor products $e_{j_1}
\otimes \cdots \otimes e_{j_n}$ with repeating factors, so that
\begin{eqnarray*}
             \hbox{Tr}\left(F_n^- \right)
             & = &
             \sum_{\stackrel{distinct}{j_1,\ldots,j_n \in J} }
             \Big< e_{j_1} \otimes \cdots \otimes e_{j_n},\ F^{\otimes n}
\Sigma_n(e_{j_1} \otimes \cdots \otimes e_{j_n})\Big>  \\
             & = &
             \sum_{\stackrel{ distinct}{j_1,\ldots,j_n \in J}}
             \left< e_{j_1} \otimes \cdots \otimes e_{j_n},\ F^{\otimes n}
(e_{j_1} \otimes \cdots \otimes e_{j_n})\right>  \\
             & \le & \hbox{Tr}\left(F^{\otimes n} \right) \ = \ 1
\end{eqnarray*}
as asserted.   \hfill $\square$

The next lemma provides a bound on the trace norm of the
remainder term $\RR_n(F)$ when $F$ is a density operator.
The bound is proportional to the {\it operator} norm of $F$.
\begin{lemma}
\label{Estimate for R}
  Let $\RR_n$ be as in (\ref{Rn}) and let $F$ be a density
operator.   Then
\begin{equation}
         \Vert\RR_n(F) \Vert _{tr} \ \le \  2n(n-1)\| V \| \ \|F\|.
\end{equation}
\end{lemma}
\noindent {\bf Proof}: \qquad    {From} (\ref{Rn}) we see that $\RR_n(F)$ equals
\[
         \Bigg\{
           \sum_{\stackrel{j,k =1}{j \ne k}}^{n}  \Big(
           V_{j,n+1} F^{\otimes n+1}U_{(k,n+1)} \ - \    F^{\otimes
n+1}U_{(k,n+1)}  V_{j,n+1}  \Big) \left( \Sigma_n \otimes I
\right)
          \Bigg\}_{:n}.
\]
Since $U_{(k,n+1)} $ commutes with $F^{\otimes n+1}$ and since
$\Sigma_n \otimes I $ commutes with $\sum_{j,k: j \ne k}
U_{(k,n+1)} V_{j,n+1} $, it follows that $\RR_n(F)$ equals
\[
        \Bigg\{  \sum_{\stackrel{j,k =1}{j \ne k}}^{n}
        V_{j,n+1} U_{(k,n+1)} \big(F_n^- \otimes F\big)
        \Bigg\}_{:n}
       \ - \
       \Bigg\{
        \big(F_n^- \otimes F\big)
       \sum_{\stackrel{j,k =1}{j \ne k}}^{n}    U_{(k,n+1)} V_{j,n+1}
       \Bigg\}_{:n}\ .
\]
Since the trace norm of a trace class operator equal the trace norm of its
adjoint, it follows that
\begin{eqnarray}
      \| \RR_n(F) \|_{tr}
      & \le &
       2 \ \Bigg\|   \sum_{\stackrel{j,k =1}{j \ne k}}^{n}
        \Big\{  V_{j,n+1} U_{(k,n+1)} \big(F_n^- \otimes F\big)
       \Big\}_{:n}
      \Bigg\|_{tr}  \nonumber \\
      & \le &
      2n(n-1) \Big\|   \Big\{  V_{n-1,n+1} U_{(n,n+1)} \big(F_n^- \otimes F\big)
       \Big\}_{:n}
      \Big\|_{tr} \ .
\label{inequality}
\end{eqnarray}
But one may verify directly that
\begin{equation}
    \Big\{  V_{n-1,n+1} U_{(n,n+1)} \big(F_n^- \otimes F\big)
       \Big\}_{:n}
    \ = \
     (I^{\otimes n-1} \otimes F)V_{n-1,n}F_n^- ,
\label{claim}
\end{equation}
so that, by (\ref{basic inequality}) and Lemma~\ref{estimate for Fn-},
\[
 \Big\|   \Big\{  V_{n-1,n+1} U_{(n,n+1)} \big(F_n^- \otimes F\big)
       \Big\}_{:n}
      \Big\|_{tr}
      \ \le \
      \|  F \| \    \| V \| \    \|F_n^- \|_{tr}  \ \le \|  F \| \    \| V \|.
\]
Substituting this in (\ref{inequality}) yields (\ref{Estimate for
R}).

To verify (\ref{claim}), choose an orthonormal basis $\{e_j\}_{j \in J}$
for
$\BbH$ and check that
the operators on both sides of (\ref{claim}) have the same matrix
elements relative to the basis $\big\{e_i \otimes e_j : \ i,j \in J \big\}$.
\hfill $\square$

Let $F(t)$ be a solution of the TDHF equation (\ref{TDHF}).     Since the
(operator) norm of $F(t)$ is constant, it follows from
Lemma~\ref{Estimate for R} that
\[
             \Vert\RR_n(F(t)) \Vert _{tr} \ \le \ 2  n(n-1) \Vert V \Vert
\ \|F(0)\|
\]
for all $ t \ge 0$.  With this estimate for $\RR_n$ and estimate
(\ref{estimates for N-particle hierarchy stuff}) for $\EE_n$, it
follows that $\hbox{Err}(t,N, n)$ of equation (\ref{err})
satisfies
\[
       \| \hbox{Err}(t,N, n) \|_{tr} \ \le \  2  n^2 \Vert V \Vert
        \left(
        \frac{2}{N}    \ + \  \|F(0)\|
        \right)
\]
and $\varepsilon(N,n,t)$ of equation (\ref{eps}) satisfies
\begin{eqnarray}
         \varepsilon(N,n,t)
         & = &
         2n^2 \|V\|  \frac{t}{\hbar}
        \left(
        \frac{2}{N}    \ + \  \|F(0)\|
        \right)
          \nonumber \\
         &    & + \
         \big\|  E_{N,n}(0) \big\|_{tr}  .
\label{estimate for eps}
\end{eqnarray}

%%%%%%%%%%%%%%%%%%%%%%%%%%%%%%%%%%%%%%%%%%%%%%%%%%%%%%%%%%%%%%%%%%%%
%
%
%%%%%%%%%%%%%%%%%%%%%%%%%%%%%%%%%%%%%%%%%%%%%%%%%%%%%%%%%%%%%%%%%%%%

\section{Proof of the theorem}

Equipped with the estimates of the preceding sections, we proceed to the proof
of  Theorem~\ref{Main result}.

So, let us assume that $D_N(0)$ is a symmetric density for each $N$ and that
the sequence $\{D_N(0)\}$ has Slater closure.   Let $D_N(t)$ be the solution
of  (\ref{N-particle Schrodinger})
with initial value $D_N(0)$, and let $F^{(N)}(t)$ be the solution of the TDHF
equation (\ref{TDHF}) whose
initial value is $F^{(N)}(0)=D_{N:1}(0)$.    Let
$\left\{F^{(N)}\right\}^-_n(t)$ denote
$\big\{F^{(N)}(t)\big\}^{\otimes n}\Sigma_n$ and let $E_{N,n}(t)$
denote the difference between $D_{N:n}(t)$ and
$\left\{F^{(N)}\right\}^-_n(t)$.

We have the upper bound (\ref{Bardos inequality}) for the trace norm of
$E_{N,n}(t)$, into which we now substitute the estimates (\ref{estimate for
eps}).  In the same stroke, we will replace the binomial coefficients
$\binom{n+k-1}{n-1}$ with the larger quantities $(n+k)^n/n!$
and we will use the fact that $\sup\limits_{s \in [0,t]}\big\{ \big\|
E_{N,n+m+1}(s) \big\|_{tr} \big\} \le  2$ by Lemma~\ref{estimate for Fn-}.
Also, let us set $T = 2\|V\|t / \hbar.$
We obtain
\begin{eqnarray}
       \big\| E_{N,n}(t) \big\|_{tr}
       & \le &
        \frac{1}{n!}  \sum_{k = 0}^{m}  (n+k)^n   \big\|
E_{N,n+k}(0)
\big\|_{tr}  T^k   \nonumber \\
       & + &
      \frac{1}{n!}
      \sum_{k = 0}^{m}  (n+k)^{n+2}
       \Big(
            \frac{2}{N}    \ + \  \big\|F^{(N)}(0)\big\|
        \Big)
        T^{k+1} \nonumber \\
        & + &
       \frac{2}{n!}  (n+m)^n T^m
\label{Bardos revisited}
\end{eqnarray}
for $m \le N-n-1$.  Fix $T$ to be less than $1$, i.e., fix $t <
\hbar/2\|V\|$.   For fixed $n$, consider the limit of the
right-hand-side of (\ref{Bardos revisited}) as $N$ and $m$ tend to
infinity. The individual terms (fixed $k$) tend to $0$, for
$\big\|F^{(N)}(0)\big\|$ tends to $0$ by Lemma~\ref{lemma1} and
$\big\|  E_{N,n+k}(0) \big\|_{tr}$ tends to $0$ thanks to the
hypothesis that  $\{D_N(0)\}$ has Slater closure (recall that
$F^{(N)}(0)=D_{N:1}(0)$).  On the other hand, the series on the
right-hand-side of  (\ref{Bardos revisited}) are dominated,
uniformly with respect to $m$, by a series that converges
absolutely for $T < 1$.
   It follows that
\begin{equation}
       \lim_{N \rightarrow \infty}   \big\| E_{N,n}(t) \big\|_{tr}   \ = \  0
\label{error tends to zero}
\end{equation}
if $t < \hbar/2\|V\|$. When $n=1$, this shows that $
       \lim\limits_{N \rightarrow \infty}   \big\|   D_{N:1}(t) - F^{(N)}(t)
\big\|_{tr}  = 0
$
and consequently
\[
       \lim_{N \rightarrow \infty}   \Big\|   D_{N:1}^{\otimes n}(t)
\Sigma_n \ - \  \big\{F^{(N)}\big\}^-_n(t) \Big\|_{tr}   \ = \  0
\]
for $n>1$ and $t < \hbar/2\|V\|$.  {From} (\ref{error tends to
zero}) again it follows that, for any $n$ and any $t <
\hbar/2\|V\|$,
\[
      \lim_{N \rightarrow \infty}   \Big\|   D_{N:n}(t) \ - \
D_{N:1}^{\otimes n}(t) \Sigma_n \Big\|_{tr}   \ = \  0,
\]
i.e.,   $\{D_N(t)\}$ has Slater closure.  This proves the theorem
up to $t = \hbar/2\|V\|$.

Let $\tau = \hbar/3\|V\|$; the previous argument shows that the
theorem holds for $t\in[0,\tau]$. At time $\tau$, it is no longer
the case that $D_{N:1}(\tau) =  F^{(N)}(\tau)$. However,
$\big\|E_{N,n+k}(\tau)\big\|_{tr}\to 0$ and
$\big\|F^{(N)}(\tau)\big\|\to 0$ as $N$ tends to infinity --- to
see this, use (\ref{nonlinearUnitary}) and the fact that
$\big\|F^{(N)}(0)\big\|\to 0$. An argument nearly identical to the
one above shows that the theorem holds for $t\in[\tau,2\tau]$.
This argument may be repeated to establish the conclusion of the
theorem on each interval of the form $[k\tau,(k+1)\tau]$ for each
nonnegative integer $k$, and hence for all $t>0$. \hfill $\square$

\section{Appendix: TDHF equations for wavefunctions}

The time-dependent Hartree-Fock equations were invented by Dirac
\cite{Dirac}.  He wrote the TDHF equations as a system of coupled
Schr\"odinger equations for $N$ time-dependent electronic
orbitals, and thence obtained the TDHF equation for the density
operator.  In this article we have studied the TDHF in the latter
form: as a von Neumann equation for the density operator.  This
appendix explains how to recast the wavefunction formulation of
the TDHF equations into the language of density operators used in
this paper.

The starting point in this discussion is the linear $N$-body
Schr\"odinger
\begin{equation}
\label{N-Schrod} i\hbar \frac{\partial}{\partial t} \Psi_N \ = \
-\frac{\hbar^2}{2} \sum_{k=1}^N\Delta_{x_k}\Psi_N
    + \frac{1}{N} \sum_{1\le k<l\le N}V(x_k-x_l)\Psi_N
\end{equation}
where $\Psi_N\equiv\Psi_N(t,x_1,\ldots,x_N)$ is the $N$-particle
wavefunction.  (Note that the interaction term has been multiplied
by $1/(N)$.  This scaling has been introduced so that $N
\longrightarrow \infty$ may yield a {\it mean-field} equation for
the single-particle density, namely, the TDHF equation.) The
dynamics defined by (\ref{N-Schrod}) is unitary on
$L^2((\BbR^3)^N)$. Therefore,
\[
 \int
\left|\Psi_N(t,x_1,\ldots,x_N)\right|^2dx_1\ldots dx_N \ = \ 1
\]
for all $t\ge 0$ if the same equality holds at $t=0$ (as is the
case if $|\Psi_N|^2$ is meant to be interpreted as the probability
density of the system of $N$ particles in its configuration
space). In the language of (\ref{Schrodinger equation for the
Intro}), $L_k=-\frac{\hbar^2}{2}\Delta_{x_k}$ while $V_{kl}$
denotes the multiplication by $V(x_k-x_l)$.  The TDHF equations
corresponding to (\ref{N-Schrod}) may be written as a system of
$N$ coupled Schr\"odinger equations for orthonormal orbitals
$\psi_1(t,x),\psi_2(t,x), \ldots,\psi_N(t,x)$:
\begin{eqnarray}
 i\hbar\frac{\partial}{\partial t} \psi_k(t,x) & = &
 -\frac{\hbar^2}{2}\Delta_x\psi_k(t,x) +
\psi_k(t,x)\frac1{N}\sum_{l=1}^N\int V(x-z)|\psi_l(t,z)|^2dz
\nonumber \\
&  & -\frac1{N}\sum_{l=1}^N\psi_l(t,x)
    \int V(x-z)\psi_k(t,z)\overline{\psi_l(t,z)}dz
\label{TDHF-wave}
\end{eqnarray}
The $N$ orbitals  remain orthonormal at all times; if
$\psi_1(t,x),\psi_2(t,x), \ldots,$ $\psi_N(t,x)$ is a solution of
(\ref{TDHF-wave}) and
\begin{eqnarray*}
\int\psi_k(0,x)\overline{\psi_l(0,x)}dx & = & \delta_{kl} \ \
\hbox{ then }
\\
\int\psi_k(t,x)\overline{\psi_l(t,x)}dx & = & \delta_{kl} \ \
\hbox{ for all }t\ge 0\,.
\end{eqnarray*}
One way to obtain the TDHF equations (\ref{TDHF-wave}) from the
linear $N$-particle Schr\"odinger equation (\ref{N-Schrod}) is to
solve a variational problem which would lead to (\ref{N-Schrod})
if unconstrained, but with the constraint that the $N$-particle
wave equation remain a Slater determinant at all times
\cite{Cances}.    This constraint is imposed for the sake of
obtaining an computationally amenable approximation to
(\ref{N-Schrod}), and it is not justified on physical grounds. In
effect, this paper proves that the constraint maintaining Slater
determinants at all times is rigorously justified {\it in the
mean-field limit}.

To see how the orbital form (\ref{TDHF-wave}) of the TDHF
equations relates to the TDHF equation (\ref{TDHF}) discussed in
this paper, we shall first rewrite (\ref{TDHF}) as an equation for
the integral kernel of a time-dependent density operator.  To do
this, we need to know how to translate the partial trace into the
language of integral operators, for equation (\ref{TDHF}) involves
a partial trace.  Let $T$ be a trace class operator on
$L^2(\BbR^m\times\BbR^n)$ having an integral kernel
$\rho(x,\xi,y,\eta)$ with $x,y\in\BbR^m$ and $\xi,\eta\in\BbR^n$.
The partial trace \[ T_{:m} \quad \hbox{ is the operator with
integral kernel} \quad
       \int \rho(x,z,y,z)dz.
\]

We may now convert the TDHF equation (\ref{TDHF}) into an
integro-differential equation for a time-dependent integral
kernel: Let $\rho\equiv\rho(t,x,y)$ be the integral kernel of the
operator $F(t)$ that appears in (\ref{TDHF}).  Then $ F_2^-(t)$
has integral kernel
$$
    \rho(t,x_1,y_1)\rho(t,x_2,y_2)-\rho(t,x_1,y_2)\rho(t,x_2,y_1)\,,
$$
while $[V,F_2^-(t)]_{:1}$ has integral kernel
$$
\int(V(x_1-z)-V(y_1-z))(\rho(t,x_1,y_1)\rho(t,z,z)-\rho(t,x_1,z)\rho(t,z,y_1))dz\,,
$$
and the TDHF equation (\ref{TDHF}) in the language of integral
kernels is
\begin{eqnarray}
 &&
 i\hbar\frac{\partial}{\partial t} \rho(t,x,y)\ =
\ -\frac{\hbar^2}{2}(\Delta_x-\Delta_y)\rho(t,x,y) \label{TDHF'}
\\
 &&
 +
\int(V(x-z)-V(y-z))(\rho(t,x,y)\rho(t,z,z)-\rho(t,x,z)\rho(t,z,y))dz\,.
\nonumber
\end{eqnarray}
It may be verified that a solution
$
      \psi_1(t,x), \ \psi_2(t,x), \ldots,\ \psi_N(t,x)
$
to the orbital form of the TDHF equations (\ref{TDHF-wave})
yields a solution $\rho(t,x,y)$ to the integro-differential
equation (\ref{TDHF'}) via
\[
       \rho(t,x,y) \ = \
       \frac1N\sum_{k=1}^N\psi_k(t,x)\overline{\psi_k(t,y)}\,.
\]

The rest of this appendix is meant to serve as a key for reading
this paper with the Schr\"odinger wave equation (\ref{N-Schrod})
in mind.

To the wavefunction $\Psi_N(t,\cdot)$ solution of the $N$-particle
Schr\"odinger equation (\ref{N-Schrod}) one associates the
operator $D_N(t)$ with integral kernel
\begin{equation}
\label{def-rhoN} \rho_N(t,x_1,\ldots,x_N,y_1,\ldots,y_N) =
\Psi_N(t,x_1,\ldots,x_N)\overline{\Psi_N(t,y_1,\ldots,y_N)}\,.
\end{equation}
The natural Hilbert space $\BbH$ in this context is
$\BbH=L^2(\BbR^3)$, and $\BbH^{\otimes N}$ is isomorphic to $
L^2((\BbR^3)^N)$ through the identification
$$
\psi_1\otimes\ldots\otimes\psi_N \ \longleftrightarrow \
   \prod_{k=1}^N\psi_k(x_k).
$$
The corresponding representation of the permutation group $\Pi_N$
is given by the formula
$$
(U_\pi\Psi)(x_1,\ldots,x_N)\ = \
\Psi(x_{\pi^{-1}(1)},\ldots,x_{\pi^{-1}(N)})
$$
for $\pi\in\Pi_N$ and $\Psi_N\in L^2((\BbR^3)^N)$. Hence the
projection $P_{\AA_N}$ is given by the formula
$$
(P_{\AA_N}\Psi_N)(x_1,\ldots,x_N)=\frac1{N!}
\sum_{\pi\in\Pi_N}\hbox{sgn}(\pi)\Psi_N(x_{\pi(1)},\ldots,x_{\pi(N)})\,.
$$
A wavefunction $\Psi$ is antisymmetric if it is in the image of
$P_{\AA_N}$, or equivalently, if
$$
\Psi(x_{\pi(1)},\ldots,x_{\pi(N)}) \ = \
\hbox{sgn}(\pi)\Psi(x_1,\ldots,x_N)
$$
for all $\pi \in \Pi_N$.  If $\Psi_N$ is antisymmetric then the
rank $1$ orthogonal projector $P_{\Psi_N}$ with integral kernel as
in (\ref{def-rhoN}) is fermionic in the sense of
(\ref{antisymmetric}).   Property (\ref{antisymmetric}) extends to
all convex combinations of such projectors $P_{\Psi_N}$, the
fermionic density operators discussed in this article. Property
(\ref{antisymmetric}) implies that fermionic density operators
commute with all of the operators $U_{\pi} $, so that the integral
kernel of a fermionic density operator is symmetric in the sense
that
$$
\rho_N(x_1,\ldots,x_N,y_1,\ldots,y_N) \ = \ \rho_N
(x_{\pi(1)},\ldots,x_{\pi(N)},y_{\pi(1)},\ldots,y_{\pi(N)}))
$$
for all $\pi\in\Pi_N$.

In the case where $\Psi_N = \psi_1\otimes\ldots \otimes\psi_N$
with $\psi_1, \ldots, \psi_N$ orthonormal, one finds that the
Slater determinant $\sqrt{N!}\ P_{\AA_N}(\Psi_N)$ truly is a
determinant:
$$
    \sqrt{N!}(P_{\AA_N}\Psi_N)(x_1,\ldots,x_N)
    \ = \ \frac1{\sqrt{N!}} \left|
           \begin{array}{cccc} \psi_1(x_1) & \psi_1(x_2) & \cdots & \psi_1(x_N) \\
                               \psi_2(x_1) & \psi_2(x_2) & \cdots & \psi_2(x_N) \\
                               \vdots      & \vdots      &        & \vdots      \\
                               \psi_N(x_1) & \psi_N(x_2) & \cdots & \psi_N(x_N) \\
           \end{array}
           \right|.
$$

\medskip

%%%%%%%%%%%%%%%%%%%%%%%%%%%%%%%%%%%%%%%%%%%%%%%%%%%%%%%%%%%%%%%%%%

\bigskip
\noindent
{\bf Acknowledgement.} {\em

This research was supported by the French-Austrian ``Amadeus"
program (\"OAD V6) and by the Austrian START project ``Nonlinear
Schr\"odinger and quantum Boltzmann equations" of N.J.M. Also,
F.G. was supported by the Institut Universitaire de France. The
first three authors thank the ESI in Vienna for its hospitality.
We also express our gratitude to Xavier Blanc, Eric Canc\`es and
Claude Le Bris for numerous and helpful discussions on this paper.
}

%%%%%%%%%%%%%%%%%%%%%%%%%%%%%%%%%%%%%%%%%%%%%%%%%%%%%%%%%%%%%%%%%%
%%%%%%%%%%%%%%%%%%%%%%%%%%%%%%%%%%%%%%%%%%%%%%%%%%%%%%%%%%%%%%%%%%

\newpage

\centerline{\bf R\'esum\'e}
\bigskip
On montre dans ce travail que les \'equations d'\'evolution de
Hartree-Fock d\'ecrivent la limite de l'\'equation de
Schr\"odinger \`a $N$ corps pour $N$ tendant vers l'infini et une
constante de couplage en $O(1/N)$ et pour des donn\'ees initiales
proches de d\'eterminants de Slater. On ne consid\`ere ici que le
cas de potentiels d'interaction binaires, sym\'etriques et
born\'es. Lorsque $N\rightarrow \infty$, on montre que la suite
des traces partielles ``\`a un corps" de l'op\'erateur densit\'e
\`a $N$ corps converge, au sens des op\'erateurs \`a trace, vers
la solution de l'\'equation de Hartree-Fock sous forme
op\'eratorielle.

\smallskip
\noindent
{\bf Mots-cl\'es:} Equations de Hartree-Fock; Probl\`eme \`a $N$ corps quantique;
        Approximation de champ moyen


\newpage

\begin{thebibliography}{X}

\bibitem{Alicki} R. Alicki and J. Messer.   Nonlinear quantum
dynamical semigroups for many-body open systems.   {\it Journal of
Statistical Physics} 32 (2): 299 - 312, 1983.

\bibitem{BGM}
C.Bardos, F. Golse and N.J. Mauser.  Weak coupling limit of the
$N$-particle Schr\"odinger equation.  {\it Mathematical Analysis
and Applications} 7 (2): 275 - 293, 2000.

\bibitem{BEGMY}
C. Bardos, L. Erd\"os, F. Golse, N.J. Mauser and H.-T. Yau.
Derivation of the Schr\"odinger-Poisson equation from the quantum
$N$-particle Coulomb problem.
{\it C. R. Acad. Sci. S\'er. I Math} 334: 515--520, (2002).


\bibitem{Boltzmann}L. Boltzmann. {\it Lectures on Gas
Theory}. Dover Publications, New York, 1995.

\bibitem{Bove}  A. Bove, G. Da Prato, and G. Fano.
An existence proof for the Hartree-Fock time-dependent problem
with bounded two-body interaction.   {\it Communications in
Mathematical Physics} 37: 183-191, 1974.

\bibitem{Cances}
E. Cances, and C. Le Bris. On the time-dependent Hartree-Fock
equations coupled with a classical nuclear dynamics.   {\it Math.
Models Methods Appl. Sci.} 9 (7): 963--990, 1999.


\bibitem{Dirac} P. Dirac, Note on exchange phenomena in the Thomas atom,
Proc. Cambridge Philosophical Society 26 (1930)  pp. 376 - 385


\bibitem{DW} N. G. Duffield and R. F. Werner.
Local Dynamics of Mean--Field Quantum Systems.  {\it Helvetia
Physica Acta} 65, 1016--1054, 1992


\bibitem{EY}
L. Erd\"os and H.-T. Yau.   Derivation of the nonlinear
Schr\"odinger equation with Coulomb potential.
{\it Preprint} (2002).


\bibitem{Thesis} A. D. Gottlieb. {\it Markov transitions and the
propagation of chaos} (PhD Thesis).   Lawrence Berkeley National
Laboratory Report, LBNL-42839, 1998.

\bibitem{Gru}F. A. Gr\"unbaum.  Propagation of chaos for the
Boltzmann equation. {\it Archive for Rational Mechanics and
Analysis}  42: 323-345, 1971.


\bibitem{Kac55}M. Kac. Foundations of kinetic theory. {\it
Proceedings of the Third Berkeley Symposium on Mathematical
Statistics and Probability, Vol III}.  University of California
Press, Berkeley, California, 1956.

\bibitem{Kac}M. Kac. {\it Probability and Related Topics in
Physical Sciences}.  American Mathematical Society, Providence,
Rhode Island, 1976.

\bibitem{McK66}H. P. McKean, Jr.  A class of Markov processes
associated with nonlinear parabolic equations.  {\it Proceedings
of the National Academy of Science} 56: 1907-1911, 1966.

\bibitem{Meleard}S. M\'el\'eard. Asymptotic behavior of some
interacting particle systems; McKean-Vlasov and Boltzmann models.
{\it Lecture Notes in Mathematics}, 1627. Springer, Berlin, 1995.

\bibitem{Sp}
H. Spohn, Kinetic Equations from Hamiltonian Dynamics. {\it Rev.
Mod. Phys.} 53: 600 -- 640, 1980.  (See Theorem 5.7)

\bibitem{Szn}A. Sznitman. \'Equations de type de Boltzmann,
spatialement homog\`enes.  {\it Zeitschrift f\"ur
Wahrscheinlichkeitstheorie und verwandte Gebiete} 66: 559-592,
1984.

\bibitem{Sznitman}A. Sznitman. Topics in propagation of chaos.
{\it Lecture Notes in Mathematics}, 1464. Springer, Berlin, 1991.




\end{thebibliography}
\end{document}